\documentclass{PoS}
\usepackage{cite,color,amsmath,amsfonts,graphicx} 
 \usepackage[utf8]{inputenc}
\newcommand{\be}{\begin{equation}}
\newcommand{\ee}{\end{equation}}
\newcommand{\gM}{\mathcal{M}}

\newcommand{\red}[1]{ \color{red} #1 \color{black} }
\newcommand{\ui}{\underline{i}} 
\newcommand{\uj}{\underline{j}}

\newcommand{\uk}{\underline{k}} 
 
\newcommand{\p}{\wedge}

\title{Orbifolds and Orientifolds as O-folds}

\ShortTitle{Orbifolds and Orientifolds as O-folds}

\author{\speaker{Chris Blair}
\\
Theoretische Natuurkunde, Vrije Universiteit Brussel, and the International Solvay Institutes,  Pleinlaan 2, B-1050 Brussels, Belgium 
\\        E-mail: \email{cblair@vub.ac.be}}

\abstract{
We consider quotients of string and M-theory by discrete subgroups of the U-duality group.
This results in what we call O-folds, which are generalisations of orbifolds and orientifolds, and generically involve non-geometric identifications between physical coordinates and dual winding coordinates. 
A simple $\mathbb{Z}_2$ quotient encodes the half-maximal duality web, describing on the same footing the Ho\v{r}ava-Witten description of M-theory on an interval, the heterotic theories, and type II in the presence of orientifold planes. This can be analysed using exceptional field theory, including the introduction of extra vector multiplets at the fixed points. 
This is an overview of the paper \cite{Blair:2018lbh}, based on a talk given at the Corfu Summer Institute 2018.}

\FullConference{Corfu Summer Institute 2018 "School and Workshops on Elementary Particle Physics and Gravity"\\
		(CORFU2018)\\
		31 August - 28 September, 2018\\
		Corfu, Greece}

\begin{document}

\tableofcontents

\section{Introduction}

The set of backgrounds on which string theory makes sense is not limited to conventional spacetime geometries of the sort first encountered in an introduction to general relativity.
We can define string theory on orbifolds \cite{Dixon:1985jw}, where the target space is a manifold quotiented by some discrete group. 
Such a geometry is singular at the fixed points of the group action, yet the string theory on such a background can be shown to be consistent (with the introduction of extra ``twisted sectors'' localised at the fixed points). 
By combining the target space quotient with additional quotients on the worldsheet (in particular, by worldsheet parity), we obtain string theory orientifolds \cite{Sagnotti:1987tw,Horava:1989vt,Dai:1989ua}, in which the fixed points of the spacetime action mark the location of orientifold planes, non-perturbative objects with negative tension. Consistency demands the inclusion of D-branes to cancel this negative charge, hence introducing open string degrees of freedom localised on the branes.
Meanwhile, a similar construction is used in the Ho\v{r}ava-Witten description of M-theory on an interval \cite{Horava:1995qa,Horava:1996ma}. We can view the interval as a circle quotiented by a $\mathbb{Z}_2$ spatial reflection, then the fixed points of the quotient are the interval endpoints, and anomaly cancellation in 11-dimensional supergravity means that there must be additional degrees of freedom ($\mathrm{E}_8$ gauge multiplets) localised at these endpoints. Shrinking the length of the interval to zero, this reduces to the $\mathrm{E}_8 \times \mathrm{E}_8$ heterotic string, and can then be related by string duality to theories involving orientifolds.

String theory is also believed to make sense on non-geometric backgrounds (for a recent comprehensive review, see \cite{Plauschinn:2018wbo}). 
These are backgrounds in which local patches are related by duality transformations, which generically mix together metric and form field components (hence not geometric).
These include T-folds, when the background is only well-defined up to transformations by the stringy T-duality symmetry, and also more generally U-folds, which are the non-perturbative generalisation making use of the full U-duality symmetry.
At fixed points in the moduli space of T- or U-fold compactifications, the gluing by duality transformations becomes a quotient \cite{Dabholkar:2002sy}, suggesting a general notion of non-geometric orbifolds and orientifolds (note that one can view asymmetric orbifolds, in which the left- and right-movers are orbifolded differently, as being of this type). 

String theory on a $d$-dimensional torus is characterised not only by momenta $p_i$ in the toroidal directions $Y^i$ but by non-trivial winding $w^i$.
We can introduce the T-dual coordinates, $\tilde Y_i$, which are conjugate to the string windings $w^i$, and parametrise a dual geometry. 
It is possible to formulate the string worldsheet theory such that the target space is the doubled geometry with coordinates $Y^M  = (Y^i,\tilde Y_i)$ transforming in the fundamental representation of the T-duality group $O(d,d)$ \cite{Duff:1989tf,Tseytlin:1990nb, Tseytlin:1990va,Siegel:1993xq, Siegel:1993th,Hull:2004in, Hull:2006va}, and use this to make sense of T-folds.
For M-theory, or string theory including D-branes and other non-perturbative objects, the analogous picture involves dual coordinates conjugate to general brane windings, filling up representations $Y^M  = (Y^i, \tilde Y_{i_1 \dots i_p}, \dots )$ of the U-duality groups $\mathrm{E}_{d(d)}$, as first investigated for the M2 brane in \cite{Duff:1990hn}.
Generic non-trivial duality transformations act to transform the ``physical'' coordinates $Y^i$ and the duals into each other.  

The duality groups $O(d,d)$ and $\mathrm{E}_{d(d)}$ are also visible in supergravity reduced to $n$ dimensions (where $n+d=10$ or $11$). 
A long-standing question has been whether these symmetries are present directly in higher dimensions prior to compactification. 
This can now be achieved using double field theory (DFT) \cite{Siegel:1993xq, Siegel:1993th, Hull:2009mi} and exceptional field theory (ExFT) \cite{Berman:2010is, Hohm:2013pua}, lifting the global $O(d,d)$ or $\mathrm{E}_{d(d)}$ acting in $n$ dimensions to a \emph{local} $O(d,d)$ or $\mathrm{E}_{d(d)}$ symmetry in an extended $n+K$ dimensional geometry with coordinates $(X^\mu, Y^M)$ (with $\mu=1,\dots,n$ and $Y^M$ in a $K$-dimensional representation of $O(d,d)$ or $\mathrm{E}_{d(d)}$). 
The extended geometry is described by metrics $g_{\mu\nu}$, $\gM_{MN}$, and a set of gauge fields, all of which may in principle depend on the full set of coordinates $Y^M$, and which transform under local symmetries including \emph{generalised diffeomorphisms} based on the groups $O(d,d)$ or $\mathrm{E}_{d(d)}$ rather than on $\mathrm{GL}(K)$.
A consistency requirement is that the true coordinate dependence is however restricted, with for instance all fields and gauge parameters demanded to depend only on at most $d$ physical coordinates $Y^i \subset Y^M$, such that the $n+K$ dimensional theory is in fact equivalent to 10- or 11-dimensional supergravity. 

The crucial advantage of DFT and ExFT is that once one has the covariant formulation in terms of the extended coordinates $Y^M$, one can take \emph{different} solutions of the consistency conditions such that different sets of coordinates are regarded as physical. In this way, one recovers different (dual) supergravity theories in 10 or 11 dimensions, all of which are contained within the DFT/ExFT framework.
For instance, ExFT (for any of the exceptional groups) can be reduced either to 11-dimensional maximal supergravity or to 10-dimensional type IIB supergravity by choosing the $d$ or $d-1$ physical $Y^i$ appropriately \cite{Hohm:2013pua,Blair:2013gqa}.
Although motivated originally by the duality groups appearing on toroidal compactification, after truncating out the unphysical dual coordinate dependence, DFT/ExFT are equivalent to the full supergravity theories in 10/11 dimensions with no reference to compactification or particular backgrounds. They therefore provide a very efficient and powerful approach to standard supergravity even before one considers, for example, U-fold backgrounds, or relaxations of the consistency constraints to obtain non-geometric compactifications.

In this contribution to the proceedings of the Corfu Summer Institute 2018, we will summarise the results of our (somewhat lengthy) paper \cite{Blair:2018lbh}.
The purpose of this paper was to study orbifolds, orientifolds and their generalisations using exceptional field theory. 
We introduced the following definition for such backgrounds: 
\begin{itemize}
\item An \textbf{O-fold} results by taking the quotient of a string/M-theory background by a discrete subgroup of the $\mathrm{E}_{d(d)}$ duality symmetry
\end{itemize}
Note the name suggests the close relationship to T- and U-folds mentioned earlier. 
We argued that a natural setting to study O-folds is exceptional field theory. We can list some generic features of O-folds in ExFT:
\begin{itemize}
\item The quotient acts \emph{geometrically} in a straightforward manner on the extended coordinates $Y^M$ and the ExFT fields, but may be highly non-geometric when viewed in terms of the standard spacetime fields.
\item Depending on how we choose the physical coordinates $Y^i \subset Y^M$, the fixed points of the quotient action may occur only in the dual directions, or only in the physical directions, or in some more complicated manner. Viewing the fixed points as defining a generalised O-plane, the intersection of this plane with the physical geometry describes different types of fixed point planes in spacetime.
\end{itemize}
A somewhat trivial, but illustrative, example of how an ExFT O-fold ``geometrises'' a spacetime quotient which is not solely geometric is provided by the Ho\v{r}ava-Witten case.
There, though we quotient the spacetime background by a geometric action (a reflection in a circle direction), the 11-dimensional three-form transforms with an additional minus sign (whereas if this was a geometric orbifold it would transform only under the same reflection), such that the Chern-Simons term in the supergravity action is invariant. When we lift to ExFT, the three-form is unified with the metric and the $\mathbb{Z}_2$ reflection that implements the Ho\v{r}ava-Witten quotient acts on the ExFT fields naturally as an $\mathrm{E}_{d(d)}$ transformation, i.e. as a (generalised) geometric transformation.

Evidently, the space of all possible O-folds is quite vast. 
We therefore focused on some particularly interesting examples. Namely:
\begin{itemize}
\item We studied only O-fold quotients which preserve half the supersymmetry, which can in fact be classified given a fixed ExFT group $\mathrm{E}_{d(d)}$.
\item We analysed in detail the simplest $\mathbb{Z}_2$ quotient which describes the ``half-maximal duality web'' and showed how in this case it is possible to introduce additional degrees of freedom at the fixed points in an $\mathrm{E}_{d(d)}$ covariant manner.
\end{itemize} 

This article has the following structure.
In section \ref{eft}, we give a very brief introduction to the ideas of exceptional field theory.
Then in section \ref{web}, we review the orbifold and orientifold construction of what we call the ``half-maximal duality web'', which links the heterotic theories, type II in the presence of orientifolds, and M-theory on an interval. 
We follow this in section \ref{O} by explaining how we define half-maximal O-folds within ExFT.
In section \ref{ex}, we focus on an example $\mathbb{Z}_2$ quotient of the $\mathrm{SL}(5)$ ExFT, and demonstrate how it recovers all of the theories in the half-maximal duality web of section \ref{web}.
Then in section \ref{loc}, we outline how to include additional gauge fields corresponding to the (possibly localised) extra degrees of freedom expected in each quotient construction.

\section{Exceptional field theory}
\label{eft} 

The ExFT construction (following \cite{Hohm:2013pua,Hohm:2013vpa}) closely resembles a supergravity theory in $n$ dimensions, but with the $n$-dimensional coordinates $X^\mu$ supplemented by a further set of coordinates $Y^M \in R_1$, where $R_1$ is a particular representation of the group $\mathrm{E}_{d(d)}$, with $n+d = 11$. 
The directions $X^\mu$ may be referred to as external. 
The fields of the theory, depending formally on all coordinates $(X^\mu, Y^M)$, include an external metric, $g_{\mu\nu}$, a generalised metric, $\gM_{MN}$, and a set of generalised gauge fields collectively known as the tensor hierarchy.
These fields may be written carrying both external and $R_1$ indices as $\mathcal{A}_\mu{}^M$, $\mathcal{B}_{\mu\nu}{}^{MN}$, $\mathcal{C}_{\mu\nu\rho}{}^{MNP}$, $\dots$. More precisely, we in fact have $\mathcal{B}_{\mu\nu} \in R_2$, $\mathcal{C}_{\mu\nu\rho} \in R_3$, $\dots$, where $R_2,R_3,\dots$ denote particular further representations of $\mathrm{E}_{d(d)}$, with for instance $R_2 \subset (R_1 \otimes R_1)_{sym}$, $R_3 \subset (R_1 \otimes R_1 \otimes R_1)$.

The local symmetry associated to infinitesimal transformations $\delta Y^M = - \Lambda^M$ of the extended coordinates is the generalised diffeomorphism symmetry, acting on generalised vectors $V^M$ via
\be
\delta_\Lambda V^M \equiv \mathcal{L}_\Lambda V^M =  \Lambda^N \partial_N V^M - \alpha P^M{}_K{}^N{}_L \partial_N \Lambda^L V^K + \lambda_V \partial_K \Lambda^K V^M\,,
\ee
which defines the generalised Lie derivative $\mathcal{L}_\Lambda$ on a generalised vector $V^M$ of weight $\lambda_V$. Here $P^M{}_K{}^N{}_L$ is the projector onto the adjoint of $\mathrm{E}_{d(d)}$, and $\alpha$ is a group-dependent constant \cite{Berman:2012vc}. 
Alternatively, one can write 
\be
\mathcal{L}_\Lambda V^M  = \Lambda^N \partial_N V^M -  V^N \partial_N \Lambda^M  + Y^{MN}{}_{KL} \partial_N \Lambda^K V^L + ( \lambda_V + \omega)  \partial_K \Lambda^K V^M \,,
\label{gldY}
\ee
where $Y^{MN}{}_{KL}$ is an invariant tensor of $\mathrm{E}_{d(d)}$,  and $\omega = -1/(n-2)$.
The generalised metric is a rank two tensor of weight zero under generalised diffeomorphisms, while the external metric is a scalar of weight $-2\omega$. 
Meanwhile the external one-form $\mathcal{A}_\mu{}^M$ transforms as a gauge field,
\be
\delta \mathcal{A}_\mu = D_\mu \Lambda^M - Y^{MN}{}_{KL} \partial_N \lambda_{\mu}{}^{KL} \,,\quad D_\mu \equiv \partial_\mu - \mathcal{L}_{\mathcal{A}_\mu} \,,
\label{deltaA}
\ee
under a combination of generalised diffeomorphisms and gauge transformations $\lambda_\mu{}^{MN} \in R_2$ associated to the field $\mathcal{B}_{\mu\nu}{}^{MN}$ at the next level of the tensor hierarchy ($\delta \mathcal{B}_{\mu\nu}{}^{MN} = 2 D_{[\mu} \lambda_{\nu]}{}^{MN} + \dots$). 

The further details of the construction will not be repeated here. The key point relevant to the O-fold construction is that we should require the Y-tensor be preserved by whatever we are quotienting by, so that generalised diffeomorphisms are preserved as a symmetry of ExFT.
This will of course be guaranteed if we only take quotients with respect to subgroups of $\mathrm{E}_{d(d)}$, but note that for $O(d,d)$ \cite{Baraglia:2013xqa, Blair:2018lbh}, the $\mathbb{Z}_2$ quotient leading to orientifolds is \emph{not} an element of $O(d,d)$, but does leave $Y^{MN}{}_{KL}$ invariant.

The coordinate dependence on the extended coordinates $Y^M$ is constrained by the section condition:
\be
\partial_M \otimes \partial_N \big|_{R_2} = 0 \,,
\label{ssc}
\ee
which ensures that the algebra of generalised diffeomorphisms closes. This condition is to be thought as being imposed on all pairs of derivatives acting either on a single field/gauge parameter or on two separate fields/gauge parameters.
A choice of physical coordinates $Y^i \subset Y^M$ such that \eqref{ssc} is satisfied will be called a solution of the section condition (SSC). 
Upon solving the section condition, we break the $\mathrm{E}_{d(d)}$ symmetry, and find that all fields decompose into $\mathrm{GL}(d)$ (or $\mathrm{GL}(d-1)$) representations with generalised diffeomorphisms reducing to ordinary diffeomorphisms plus gauge transformations of the spacetime $p$-forms.

Given the above fields, we can use invariance under the local symmetries -- generalised diffeomorphisms plus external diffeomorphisms and gauge transformations -- to construct (in some cases pseudo-)actions which encode the dynamics of 11- and 10-dimensional maximal supergravity together in an $\mathrm{E}_{d(d)}$ covariant fashion.

\section{The half-maximal duality web}
\label{web} 

\subsection*{Type I}

The type I superstring is obtained as the quotient Type I = Type IIB / $\Omega$, where $\Omega$ is the worldsheet parity transformation $\Omega: \sigma \rightarrow - \sigma$.
We can view this as being equivalent to adding a spacetime filling orientifold plane, called an O9 plane.
This has negative RR tension, and so we are led to add 16 D9 branes. 
The unoriented open strings ending on these branes then provide a $\mathrm{SO}(32)$ gauge sector, which is exactly what is needed to cancel the anomaly of the type I theory.

The action of the transformation $\Omega$ on the bosonic massless (SUGRA) fields of type IIB is $\Omega: (g,B_2,\phi, C_0, C_2, C_4) \rightarrow (g, - B_2,\phi, - C_0, C_2 , - C_4)$. 
Quotienting out, and including the gauge fields from the open string sector, we obtain the bosonic field content of type I SUGRA, namely: $(g,C_2, \phi, A^\alpha)$.
In the presence of the gauge fields, the Bianchi identity for the field strength $F_3$ of $C_2$ is modified, such that $dF_3 \sim \mathrm{tr} ( F \wedge F )$, where $F$ denotes the field strength of the $\mathrm{SO}(32)$ gauge field. Similarly one finds that the two-form $C_2$ transforms under gauge transformations of the $A^\alpha$, schematically $\delta C_2 \sim \mathrm{tr} ( \Lambda d A )$, for $\delta A^\alpha = D \Lambda^\alpha$. This, together with additional gravitational contributions to the gauge transformations and Bianchi identities, is vital for anomaly cancellation.

\subsection*{Type I${}^\prime$}

We can view the worldsheet parity transformation as interchanging left- and right-movers, $\Omega: X_L \leftrightarrow X_R$.
Now, T-duality acts such that $X = X_L + X_R \rightarrow \tilde X = X_L - X_R$. 
Combining $\Omega$ with T-duality we find that the T-dual of the IIB/$\Omega$ quotient is IIA / $\Omega \,\mathcal{I}$ where $\mathcal{I}: \tilde X \rightarrow  - \tilde X$.
This now acts also on the target space, as a reflection in the IIA dual circle.
We thus have in spacetime an $S_{\tilde R}^1 / \mathbb{Z}_2$ orbifold, which when combined with the $\Omega$ quotient gives an orientifold.
The fixed points are $\tilde X = 0$, $\tilde X = \pi \tilde R$, and at these fixed points we have O8 planes. 
We are now led to add 8 D8s at the fixed points, giving an $\mathrm{SO}(16) \times \mathrm{SO}(16)$ gauge sector

Acting with further T-dualities produces theories with O$p$ planes at the fixed points of a transverse $T^{9-p} / \mathbb{Z}_2$ orbifold. 

\subsection*{Ho\v{r}ava-Witten / heterotic $\mathrm{E}_8 \times \mathrm{E}_8$}

The Ho\v{r}ava-Witten description of M-theory on interval is as follows.
We view the interval as $I = S^1_R / \mathbb{Z}_2$, where the $\mathbb{Z}_2$ transformations acts as $\mathcal{I} : z \rightarrow - z$. The fixed points are again $z = 0 , \pi R$. 
The action of $\mathcal{I}$ on the 11-dimensional SUGRA fields is non-trivial: so that the 11-dimensional action is invariant, the three-form must transform with an extra minus sign, i.e. we have $\mathcal{I}: (g, C_3) \rightarrow (g,-C_3)$. 
Decomposing the 11-dimensional coordinate index $\hat \mu = (\mu,z)$, with $\mu$ 10-dimensional, we have that $g_{\mu\nu}, g_{zz}$ and $C_{\mu \nu z}$ are therefore even under $\mathcal{I}$, while $g_{\mu z}$ and $\mathcal{C}_{\mu\nu\rho}$ are odd, transforming as
\be
\mathcal{I}: g_{\mu z} (x,z) \rightarrow - g_{\mu z} (x,-z) \,\quad C_{\mu\nu\rho}(x,z) \rightarrow - C_{\mu\nu\rho} (x,-z)
\ee	
and therefore projected out at the fixed points. 
Anomaly cancellation requires additional degrees of freedom at these fixed points (which we view as 10-dimensional ``end-of-the-world'' branes), in particular we need an $\mathrm{E}_8$ super-Yang-Mills theory at each fixed point.
The Bianchi identity for the field strength $F_4$ of the three-form is then $d F_4 \neq 0$, but receives localised contributions from the field strengths of the Yang-Mills fields (plus also gravitational contributions which are also localised at the fixed point planes); similarly the even components of the three-form transform under the gauge transformations of the Yang-Mills fields, which are also of course localised, schematically $\delta C_{\mu\nu z} \sim \delta(z)\,\mathrm{tr}( \Lambda \partial_{[\mu} A_{\nu]})$.

For small $R$, M-theory on $S^1_R / \mathbb{Z}_2$ reduces to the heterotic $\mathrm{E}_8 \times \mathrm{E}_8$ string theory.
We can see this at the level of supergravity by noting that at fixed points we obtain the heterotic massless fields $(g,B_2,\phi)$ from the even components of the 11-dimensional metric and three-form:
\be
( g_{\mu\nu}^{\mathrm{het}}, B_{\mu\nu} , e^\phi ) \sim ( g_{\mu\nu}, C_{\mu\nu z} , g_{zz} ) 
\ee
with the odd components projected out. The Yang-Mills multiplets then descend to those of the $\mathrm{E}_8 \times \mathrm{E}_8$ heterotic SUGRA in the limit $R \rightarrow 0$. The two-form field $B_{\mu\nu}$ then comes with modified gauge transformations and Bianchi identity, inherited from the localised modifications of the parent three-form.

Note that in the bulk, we still have a theory described locally by maximal 11-dimensional SUGRA, with the Yang-Mills sector localised at the boundary where half the maximal degrees of freedom drop out.

\subsection*{Heterotic $\mathrm{SO}(32)$}

The other heterotic theory, that with gauge group $\mathrm{SO}(32)$ can be obtained as the strong coupling limit of the type I theory.
The two heterotic theories are T-dual to each other in the presence of Wilson lines which break the gauge groups.

\section{Orbifolds, orientifolds and O-folds}
\label{O}

\subsection{Orbifolds and orientifolds}

An orbifold is obtained as a quotient of a manifold $M$ by a discrete group $G_{\textrm{discrete}}$.
If we consider strings on an orbifold, and take an additional quotient by worldsheet parity, then we obtain an orientifold.
At fixed points of a stringy orbifold or orientifold, we will generically have additional degrees of freedom (twisted sectors), which will be needed for consistency.
For instance, for orbifolds we require modular invariance, which can be used to work out what the twisted sectors must be, while for orientifolds they may be determined by charge or anomaly cancellation.

\subsection{O-folds in ExFT}

We define O-folds to be the generalisation of orbifolds and orientifolds obtained by quotienting string or M-theory by discrete subgroups of the U-duality group in $d$ dimensions.
Quotients with respect to discrete subgroups of the U-duality have of course been studied in the literature, see for instance \cite{Dasgupta:1996ij} which suggested the term ``generalised orientifold'' for such constructions. 
We propose to study these systematically in the context of exceptional field theory, as this provides a natural setting in which the action of U-duality can be geometrised. For this reason we use the name ``O-fold'' to emphasise the similarity with T- and U-folds.

Given $Z^M{}_N \in G_{\textrm{discrete}} \subset E_{d(d)}$, we impose a quotient via an identification on the extended coordinates 
\be
(X^\mu, Y^M ) \sim (X^\mu , Y^{\prime M} ) = ( X^\mu , Z^M{}_N Y^N ) \,,
\ee
and generalised fields e.g.
\be\begin{split} 
g_{\mu\nu} ( X, Y) &\sim g_{\mu \nu} ( X, Y^\prime ) \,,\\
\gM_{MN} ( X,Y) &\sim (Z^{-1})^P{}_M (Z^{-1})^Q{}_N \gM_{PQ} (X, Y^\prime ) \,,\\
\mathcal{A}_\mu{}^M(X,Y) &\sim Z^M{}_N \mathcal{A}_\mu{}^N(X,Y^\prime) \,,
\end{split}
\ee
plus similar transformations of the other tensor hierarchy fields according to their representations.

Let us consider the coordinate identification, $Y^M \sim Z^M{}_N Y^N$.
Depending on the choice of physical coordinates in a given SSC, we may have a geometric quotient in spacetime, i.e. $Y^i \sim Y^j$, leading to conventional orbifolds/orientifolds or a non-geometric identification $Y^i \sim \tilde Y_{i j}$, which corresponds to asymmetric orbifolds and ``non-perturbative'' generalisations.
Fixed points occur when $Y^M = Z^M{}_N Y^N$.
Again, depending on the SSC, these fixed points may occur only in dual directions, or only in physical directions, or in some complicated fashion where one can not disentangle the physical and dual directions at all.
This general picture is compatible with T/U-fold compactifications, for which O-folds would appear at fixed points in moduli space \cite{Dabholkar:2002sy}.

\subsection{Half-maximal O-folds}

The standard ExFT construction leads to a theory with maximal supersymmetry \cite{Godazgar:2014nqa}. 
However, one can formulate conditions to encode backgrounds and theories with less supersymmetry in $\mathrm{E}_{d(d)}$ covariant language - in fact one can do this while working exclusively with the bosonic sector of the theory, as we will do here.
We require the existence of an $\mathrm{E}_{d(d)}$ ExFT half-maximal structure \cite{Malek:2017njj}, which for $d \leq 6$ consists of the generalised tensors (the $d=7$ case is not quite the same, and is also discussed in \cite{Malek:2017njj})
\be
J_u{}^M \in R_1 \,,\quad \hat K_{MN} \in R_{n-4} \cong \bar R_2
\ee
which are to be viewed as nowhere vanishing sections of bundles (over the physical spacetime manifold) whose fibres are the $\mathrm{E}_{d(d)}$ tensor hierarchy representations $R_1$ and $R_{n-4}$, and which obey certain compatibility conditions. Here, $u = 1,\dots d-1$ and $n=11-d$.
The existence of these objects is equivalent to the presence of spinor bilinears built out of globally defined Killing spinors, as mentioned above we will work solely with the bosonic part of the theory.

The existence of this half-maximal structure reduces the structure group to $\mathrm{Spin}(d-1) \subset \mathrm{E}_{d(d)}$ which preserves the half-maximal structure.
It follows that we can consider quotients by discrete subgroups of $\mathrm{Spin}(d-1) \subset \mathrm{E}_{d(d)}$ in order to obtain half-maximal generalised orbifolds or O-folds

\section{Example: $\mathrm{SL}(5)$ and the half-maximal duality web}
\label{ex} 

\subsection{$\mathrm{SL}(5)$ ExFT and half-maximal structure}

The $\mathrm{SL}(5)$ ExFT \cite{Berman:2010is, Berman:2011cg, Musaev:2015ces} has coordinates $(X^\mu, Y^M)$, where $\mu$ is a seven-dimensional index, and the extended coordinates $Y^M$ transform in the ten-dimensional antisymmetric representation of $\mathrm{SL}(5)$.
Letting $a,b=1,\dots,5$ denote fundamental indices of $\mathrm{SL}(5)$, we can write the coordinates carrying a pair of antisymmetric fundamental indices, $Y^M \equiv Y^{ab} = - Y^{ba} \in \mathbf{10}$.
The totally antisymmetric invariants are denoted $\eta_{abcde}$ and $\eta^{abcde}$, and the Y-tensor defining the generalised Lie derivative \eqref{gldY} is $Y^{ab,cd}{}_{ef,gh} = \eta^{abcd i} \eta_{efgh i}$, so that the section condition is \cite{Berman:2011cg}
\be
\partial_{[ab} \otimes \partial_{cd]} = 0 \,,
\label{5sc}
\ee
acting on fields/gauge parameters and on pairs of fields/gauge parameters.
This admits four-dimensional solutions, corresponding to reductions to 11-dimensional supergravity, and three-dimensional solutions, corresponding to reductions to IIA and IIB supergravity (the former is naturally contained within the M-theory solution by imposing an additional isometry, while the latter is inequivalent under $\mathrm{SL}(5)$ transformations) \cite{Blair:2013gqa}.

The generalised metric $\gM_{ab,cd}$ decomposes as $\gM_{ab,cd} = 2m_{a[c} m_{d]b}$ where $m_{ab}$ is a symmetric unit determinant matrix, parametrising $\mathrm{SL}(5) / \mathrm{SO}(5)$. The other ExFT fields are the external metric, $g_{\mu\nu}$, and the tensor hierarchy fields. 
The relevant representations are $R_1 = \mathbf{10}$, $R_2 = \mathbf{\bar{5}}$, $R_3 = \mathbf{5}$ and $R_4 = \mathbf{\bar{10}}$, so we write the fields as
$\mathcal{A}_{\mu}{}^{ab}$, $\mathcal{B}_{\mu\nu a}, \mathcal{C}_{\mu\nu\rho}{}^a$ and $\mathcal{D}_{\mu\nu\rho\sigma ab}$. (Alternatively we can write these as carrying $R_1$ indices, for instance $\mathcal{B}_{\mu\nu}{}^{MN}$ is given by $\mathcal{B}_{\mu\nu}{}^{ab,cd} = \eta^{abcde} \mathcal{B}_{\mu\nu e}$.)

The half-maximal structure consists of three generalised vectors, $J_u{}^{ab}$, $u=1,2,3$ and a generalised tensor $\hat K^a$ in the $\mathbf{5}$ of $\mathrm{SL}(5)$.
These obey \cite{Malek:2017njj}
\be
\eta_{abcde} J_u{}^{bc} J_v{}^{de} = \frac{1}{3} \delta_{uv} \eta_{abcde} J_w{}^{bc} J^{w de} \,,\quad
\eta_{abcde} \hat K^a J_u{}^{bc} J_v{}^{de} > 0 \,.
\ee
The stabiliser of this half-maximal structure can be shown to be $\mathrm{SU}(2) \subset \mathrm{SL}(5)$. Half-maximal O-folds therefore result from quotienting by discrete subgroups of $\mathrm{SU}(2)$, which admit an ADE classification. 
These subgroups can be constructed explicitly after fixing (without loss of generality) $\hat K^a = ( 0,0,0,0,\kappa)$, $\kappa \neq 0$, corresponding to splitting the index $a= (i,s)$ with $i=1,\dots,4$ (that is, $\hat K^i = 0, \hat K^s \neq 0)$.
After making this choice, one can construct the half-maximal structure and its stabiliser in terms of the self-dual and anti-self-dual 't Hooft symbols $\eta_{u,ij}$ and $\bar\eta_{u,ij}$, finding the most general $\mathrm{SL}(5)$ element which leaves the half-maximal structure invariant is of the form
\be
Z^a{}_b = \begin{pmatrix} 
\cos \frac{\theta}{2} \delta^i{}_j + \sin \frac{\theta}{2} \bar \eta_{u , ij} \frac{\theta_u}{\theta} & 0 \\ 0 & 1
\end{pmatrix}\,,\quad \theta \equiv \sqrt{ \theta_u \theta_v \delta^{uv} } \,,
\ee
involving three real parameters $\theta_u$. 
We really only want to consider stabilisers corresponding to discrete subgroups: this means restricting the parameters $\theta_u$ in order to reproduce the ADE subgroups.
For example, $\theta_3 = \frac{4\pi}{k+1}$, $\theta_1 = \theta_2 = 0$ gives the generators of the $A_k$ series, $k \geq 1$, corresponding to $\mathbb{Z}_{k+1}$ subgroups. 

Note that one could choose an M-theory solution of the section condition such that the physical coordinates are $Y^{i s}$ and the duals are $Y^{ij}$.
Then the quotient by discrete subgroups generated by the above stabiliser acts entirely in the physical spacetime, corresponding to M-theory on $\mathbb{C}^2 / \Gamma$ where $\Gamma$ are the ADE subgroups. In other choices of the section condition solution, the quotient will generically involve identifications between physical coordinates and duals, and so be non-geometric in nature.
An exception, which we can analyse in detail, corresponds to the simplest possible quotient, namely that by $\mathbb{Z}_2$.

\subsection{$\mathbb{Z}_2$ O-fold and the half-maximal duality web}

We will now focus in detail on this $\mathbb{Z}_2$, which is generated by		
\be
Z^a{}_b = \mathrm{diag}(-1,-1,-1,-1,\red{+}1) \,.
\label{Z}
\ee
This transformation acts directly on the fundamental representation, such that a single direction in the $\mathbf{5}$ is even (we emphasise this by colouring the plus sign red) and four are odd. 
On the antisymmetric representation, corresponding to the extended coordinates, we have $Y^{ab} \rightarrow Z^a{}_c Z^b{}_d Y^{cd}$.
Hence six of the coordinates $Y^{ab}$ are even under the $\mathbb{Z}_2$ and four are odd.
We quotient by the action of \eqref{Z} on both the coordinates and the ExFT fields. 
This defines the $\mathbb{Z}_2$ O-fold without making reference to a specific solution of the section condition.
We can then proceed to consider different SSCs, for each of which the $\mathbb{Z}_2$ transformation will result in different quotients in spacetime.
This means making use of the dictionary developed in \cite{Blair:2018lbh} to relate the components of the ExFT fields to components of the standard supergravity fields, allowing us to identify how the latter transform under \eqref{Z}. We will now summarise the results of this procedure. 

Note that below we will identify the $\mathrm{E}_8 \times \mathrm{E}_8$ and $\mathrm{SO}(32)$ heterotic theories according to how they appear according to their known duality relationships, however strictly speaking we do not yet have an intrinsic ExFT method for determining the gauge group.
We will discuss in the next section how one can, despite this, introduce localised vector multiplets for different gauge groups.

Our notation here is such that we first list the parity of the five-dimensional index $a$ under the transformation \eqref{Z}, then that of the coordinates $Y^M$, grouped according to whether they are physical or associated to particular brane windings and hence dual.

\subsubsection*{M-theory SSCs} 

To describe the four-dimensional M-theory solution to the section condition \eqref{5sc}, we decompose the fundamental index $a=(i,5)$, with $i=1,\dots,4$.
The coordinates split as $Y^{ab} = ( Y^{i5}, Y^{ij})$ and we impose $\partial_{ij} = 0$ acting on all fields and gauge parameters, so that $Y^{i5}$ become the physical coordinates (alongside the $X^\mu$).

\begin{itemize}
\item {Ho\v{r}ava-Witten:}{} $a=(i,5)$ parity $(\red{+}----)$
\be
\begin{array}{rcl} 
 \text{physical:} & Y^{i5} &  -+++    \\
	\text{dual (M2 $w$):} & Y^{ij} & ---+++  
\end{array} 
\label{MsecHW}
\ee
We have a reflection in one spacetime direction, $Y^{15} \rightarrow - Y^{15}$. This gives the Ho\v{r}ava-Witten interval.
We can decompose the fields to find that for example $\mathcal{C}_{\mu\nu \rho}{}^5$ is odd: this corresponds to the components $C_{\mu\nu\rho}$ of the three-form, hence the $\mathrm{SL}(5)$ $\mathbb{Z}_2$ quotient \eqref{Z} gives not only a spacetime reflection but also acts as $C_3 \rightarrow - C_3$.

\vspace{1em}
\item {Geometric orbifold:}{} $a=(i,5)$ parity $(----\red{+})$
\be
\begin{array}{rcl} 
\text{physical:} &  Y^{i5} &  ----   \\
	\text{dual (M2 $w$):} & Y^{ij} & ++++++  
\end{array} 
\label{MsecO6}
\ee
Here we find that we have reflections in four directions in spacetime, but in fact there is no action on the fields, so this is a geometric orbifold, corresponding to M-theory on $T^4/\mathbb{Z}_2$ (or $\mathbb{R}^4 / \mathbb{Z}_2$).

\end{itemize}

\subsubsection*{IIA SSCs} 

To describe the three-dimensional IIA solution to the section condition \eqref{5sc}, we decompose the fundamental index $a=(i,4,5)$, with $i=1,\dots,3$.
The coordinates split as $Y^{ab} = ( Y^{i5}, Y^{45}, Y^{ij}, Y^{i4} )$, with $\partial_{45} = 0 = \partial_{ij} = \partial_{i4}$, so that $Y^{i5}$ are the physical coordinates alongside the $X^\mu$.
We now have three types of SSC depending on whether the even direction singled out by the $\mathbb{Z}_2$ generator \eqref{Z} is one of the $i$, $4$ or $5$.

\begin{itemize}
\item {Heterotic ``$\mathrm{E}_8 \times \mathrm{E}_8$'':}{} $a = (i,4,5)$ with parity $(---\red{+}-)$
\be
\begin{array}{rcl}
\text{physical:} & Y^{i5} & +++   \\
\text{M-theory:} &  Y^{45} & -  \\
	\text{dual (F1 $w$):} & Y^{ij} & +++   \\
	\text{dual (D2 $w$):} & Y^{i4} & ---  \\ 
\end{array}
\label{AsecHet}
\ee
There are no identifications in the physical spacetime: hence the full 10-dimensional spacetime is a fixed point. Studying the action on the fields, we find that $C_1$ and $C_3$ are projected out.

\vspace{1em}
\item {Type I${}^\prime$ (O8 planes):}{} $a=(i,4,5)$ with parity $(\red{+}----)$
\be
\begin{array}{rcl}
\text{physical:} & Y^{i5} & -++   \\
\text{M-theory:} & Y^{45} & +  \\
	\text{dual (F1 $w$):} & Y^{ij} & --+  \\
	\text{dual (D2 $w$):} & Y^{i4} & -++  \\ 
\end{array}
\label{AsecO8}
\ee
We now have a reflection in one spacetime direction, $Y^{15} \rightarrow - Y^{15}$. 
The fields transform  as $(g,B_2, \phi, C_1,C_3 ) \rightarrow (g, -B_2, \phi, C_1, -C_3)$.
This agrees exactly with what happens in type IIA in the presence of O8 planes, localised at the fixed points of the $Y^{15}$ reflection.
\vspace{1em}

\end{itemize}

\begin{itemize}
\item {IIA with O6 planes:}{} $a=(i,4,5)$ with parity $(----\red{+})$ 
\be
\begin{array}{rcl}
\text{physical:} & Y^{i5} & ---   \\
\text{M-theory:} & Y^{45} & -  \\
	\text{dual (F1 $w$):} & Y^{ij} & +++  \\
	\text{dual (D2 $w$):} & Y^{i4} & +++   
\end{array}
\label{AsecO8}
\ee
We have reflections in three directions in spacetime, $Y^{i5} \rightarrow - Y^{i5}$. The fields turn out to transform as $(g,B_2, \phi, C_1,C_3 ) \rightarrow (g, -B_2, \phi, -C_1,C_3)$, matching the description of IIA in the presence of O6 planes.

\end{itemize}

\subsubsection*{IIB SSCs}{} 
To describe the three-dimensional IIA solution to the section condition \eqref{5sc}, we decompose the fundamental index $a=(i,\alpha)$, with $i=1,\dots,3$ and $\alpha=1,2$ associated to the unbroken $\mathrm{SL}(2)$ S-duality symmetry. The coordinates split as $Y^{ab} = (Y^{ij}, Y^{i \alpha}, Y^{\alpha \beta})$, and we take $\partial_{i\alpha} = 0  = \partial_{\alpha \beta}$, so the physical coordinates are the three $Y^{ij}$ (which can be dualised into a form with one index, albeit lowered). There are two types of SSC, depending on whether the even direction of the $\mathbf{5}$ singled out by the $\mathbb{Z}_2$ transformation \eqref{Z} is one of the physical directions $i$ or one of the S-duality directions $\alpha$.
\begin{itemize}

\item {Heterotic ``$\mathrm{SO}(32)$'':}{} $a = (i , {\alpha} )$ with parity $(---\red{+}-)$
\be
\begin{array}{rcl }
\text{physical:} & Y^{ij} &  +++   \\
	\text{dual (D1/F1 $w$):} & Y^{i \alpha} & \left\{\begin{array}{c}  ---  \\   +++  \end{array}\right.   \\
		\text{dual (D3 $w$):} &Y^{{\alpha} {\beta}} & - 
\end{array} 
\label{BsecHet}
\ee
Here there are no reflections in the physical directions, so we have a ten-dimensional spacetime. 
It turns out that the fields $C_0, C_2, C_4$ are projected out.

\vspace{1em}
\item {Type I:}{} $a = (i , {\alpha} )$ with parity $(----\red{+})$
\be
\begin{array}{rcl }
\text{physical:} & Y^{ij} &  +++   \\
	\text{dual (D1/F1 $w$):} & Y^{i \alpha} & \left\{\begin{array}{c} +++  \\   ---  \end{array}\right.   \\
		\text{dual (D3 $w$):} &Y^{{\alpha} {\beta}} & - 
\end{array} 
\label{BsecHet}
\ee
This is the same as the previous case but with the even S-duality direction interchanged with the odd S-duality direction.
Again there are no reflections in the physical directions, so we have a ten-dimensional spacetime, and now the fields $C_0, B_2, C_4$.
This therefore naturally corresponds to the type I theory, which is S-dual to the heterotic $\mathrm{SO}(32)$.

\end{itemize}

\begin{itemize}

\item {IIB with O7 planes:}{} $a = (i , {\alpha} )$ with parity $(\red{+}----)$
\be
\begin{array}{rcl }
\text{physical:} & Y^{ij} &  --+   \\
	\text{dual (D1/F1 $w$):} & Y^{i \alpha} & \left\{\begin{array}{c} ++-  \\   ++-  \end{array}\right.   \\
		\text{dual (D3 $w$):} &Y^{{\alpha} {\beta}} & + 
\end{array} 
\label{BsecHet}
\ee
We have reflections in two directions in spacetime, and the fields $B_2, C_2$ turn out to be odd. This corresponds to IIB in the presence of O7 planes. 
\end{itemize}

\subsection{Generalised O-planes}

We can view the above construction geometrically as follows.
As four of the coordinates $Y^M$ are odd, we have $2^4$ fixed points of the $\mathbb{Z}_2$ quotient.
Each of these fixed points gives a $7+6$ dimensional O-fold plane.
These O-fold planes can intersect with the $7+4$ or $7+3$ dimensional physical spacetime in a variety of ways.
When the odd coordinates are all dual, the fixed points do not occur in the physical directions and so the O-fold planes fill the entire physical spacetime.
Then, we obtain true 10-dimensional theories corresponding to type IIB in the presence of O9 planes (i.e. type I), or the heterotic theories (which can perhaps be associated to certain NS9 planes \cite{Bergshoeff:1998re}). 
Alternatively, there can be genuine fixed points in some of the physical directions $Y^i$, so that the O-fold plane is not spacetime filling but rather becomes for example an ordinary orientifold plane (perhaps automatically accompanied by D-branes) in spacetime, or some even more non-perturbative generalisation (in the M-theory geometric orbifold, the fixed point planes can be thought of as the strong coupling limits of O6 plus D6 brane configurations).

\section{Localised vector multiplets in the $\mathbb{Z}_2$ O-fold}
\label{loc}

We will now sketch how one can introduce localised vector multiplets, leaving most of the details to \cite{Blair:2018lbh} (see also \cite{Malek:2016bpu,Malek:2016vsh,Malek:2017njj}).
The idea is that we will study the theory at a fixed point, expanding all ExFT fields in terms of a basis of generalised tensors which do not vanish at the fixed points. We then extend this basis to include additional vector fields which are localised at the fixed points. For simplicity, let us imagine we are working with only one fixed point - letting $\mathbf{y}$ denote the collection of all (physical) odd coordinates, we will focus on the fixed point at $\mathbf{y} = 0$ (the generalisation to treat multiple fixed points simultaneously is straightforward but notationally cumbersome).
We denote the basis we need by 
\be
\omega_A{}^M = ( \omega_{\uk}{}^M , \omega^{\uk}{}^M , \omega_{\alpha}{}^M )  \,,\quad n \in R_2 \,,\quad \hat n \in R_{n-4} \,,
\label{basis}
\ee
such that for instance a generalised vector is expanded as $V^M(X,Y) = \omega_A{}^M (Y) V^A(X,Y) + \dots$, where the ellipsis denotes components that vanish at the fixed points. (Away from the fixed points we retain locally the full ExFT degrees of freedom, subject to the overall $\mathbb{Z}_2$ identification: the question of interest is to work out happens at the fixed points themselves. For this reason we will drop the ``odd'' components vanishing at the fixed points in the following discussion.) Thus at the fixed points we have generalised vectors $V^A = (V^{\uk}, V_{\uk} , V^\alpha)$, where $\uk = 1,2,3$ is used to label the six even components coming from the original generalised vector $V^M$, and $\alpha = 1, \dots , \dim G$ labels adjoint indices of some Lie group $G$, and are carried by the additional components $V^\alpha$ which only appear at the fixed point. In the basis \eqref{basis}, we therefore take $\omega_\alpha{}^M$ to be localised at the fixed point. The $\omega_A$ can be thought of as providing a basis for an enlarged (localised) tangent bundle similar to heterotic generalised geometry \cite{Coimbra:2014qaa}. (Note that $G$ is introduced ``by hand'' in this procedure, as we do not yet have an understanding of how to fix this in ExFT. In principle, this gauge group $G$ could include a Lorentz factor as used in similar circumstances in \cite{Bedoya:2014pma,Coimbra:2014qaa} so as to describe both gauge and gravitational contributions to the modified Bianchi identities.)

Then for instance the half-maximal structure is expanded as:
\be
\begin{split}
J_u{}^M(X,Y) & = J_u{}^A (X,Y) \omega_A{}^M (Y) + \dots 
\,,\\
\hat K(X,Y) & = e^{-2d(X,Y) } \hat n(Y) + \dots 
\end{split}
\ee
(the quantity $e^{-2d}$ here is related to the dilaton in the theory at the fixed points)
and the fields $\mathcal{A}$, $\mathcal{B}$ as:
\be
\begin{split}
\mathcal{A}_\mu{}^M(X,Y) & = \mathcal{A}_\mu{}^A (X,Y) \omega_A{}^M (Y) + \dots \\
\mathcal{B}_{\mu\nu}(X,Y) & = B_{\mu\nu} (X,Y)n (Y) + \dots
\end{split} 
\ee
The even basis obeys consistency conditions, including 
\be
\omega_A \p \omega_B = \eta_{AB} n
	\,,\quad
\mathcal{L}_{\omega_A } \omega_B = - f_{AB}{}^C \omega_C \,,
\ee
where $\p: R_1 \otimes R_1 \rightarrow R_{2}$ is a generalised wedge product \cite{Wang:2015hca}.
The quantities $\eta_{AB}$ and $f_{AB}{}^C$ encode the structure of the theory at the fixed points.
The latter encodes the structure constants $f_{\alpha \beta}{}^\gamma$ of the Lie group $G$, and has all other components vanishing.
The symmetric matrix $\eta_{AB}$ is written as 
\be
\eta_{AB} = \begin{pmatrix}
0 & \delta_{\ui}{}^{\uj} & 0 \\
\delta^{\ui}{}_{\uj} & 0 & 0 \\
0 & 0 & \delta(\mathbf{y}) \kappa_{\alpha \beta}
\end{pmatrix} \,,
\label{eta}
\ee
where $\kappa_{\alpha \beta}$ is a Killing form for the Lie group $G$. 
Using the above relations, we can analyse the modification to the generalised Lie derivative at the fixed points owing to the expansion in terms of the $\omega_A{}^M$.
The result is
\be
\mathcal{L}_\Lambda V^A = \Lambda^B \partial_B V^A - V^B \partial_B  \Lambda^A + \eta^{AB} \eta_{CD} \partial_B \Lambda^C V^D + f_{BC}{}^A \Lambda^B V^C \,.
\label{modgld}
\ee
Here $\partial_A = \omega_A{}^M \partial_M$, and we always take $\partial_\alpha = 0$ (this means that the components $\eta^{\alpha \beta}$ of the inverse $\eta^{AB}$, which would be problematic owing to the delta function in \eqref{eta}, do not appear). 
This generalised Lie derivative essentially coincides with that of $O(D,D+\dim G)$ heterotic DFT \cite{Siegel:1993xq, Siegel:1993th, Hohm:2011ex,Grana:2012rr,Bedoya:2014pma,Coimbra:2014qaa}, but remarkably it has inherited from the ExFT starting point the ability to describe not just the heterotic theories but the other theories present in the half-maximal duality (in particular the delta function in \eqref{eta} allows us to pick out theories with fixed points in spacetime).

For example, we can consider the gauge transformations of the external one-form, $\mathcal{A}_\mu{}^M$, as in equation \eqref{deltaA}.
Consider first the expansion of both $\mathcal{A}_\mu{}^M$ and the gauge parameters $\Lambda^M$ in the basis \eqref{basis}. The components of the one-form are then $\mathcal{A}_\mu{}^A = (  A_\mu{}^{\uk}  ,   {A}_{\mu \uk} , \tilde {A}_\mu{}^{\alpha})$. The quantity ${A}_\mu{}^{\uk}$ is related to components of the metric in the theory restricted to the fixed points, while ${A}_{\mu \uk}$ correspond to non-vanishing components, with one external index, of some form field (see below for some examples). Meanwhile $\tilde {A}_\mu{}^\alpha$ are interpreted as gauge fields localised at the fixed points. 
Similarly the components $\Lambda^A = ( \Lambda^{\uk} , \Lambda_{\uk} , \tilde \Lambda^\alpha )$ correspond to diffeomorphisms in the directions of the fixed plane, form-field gauge transformations, and gauge transformations of the $\tilde{A}_\mu{}^\alpha$, respectively. In fact, the form of the modified Lie derivative implies that it is natural to ``twist'' these components such that $A_{\mu \uk} \rightarrow A_{\mu \uk} + \sigma \delta(\mathbf{y}) \mathrm{tr}\,( \tilde A_\mu \tilde A_{\uk} )$, $\Lambda_{\uk} \rightarrow \Lambda_{\uk} + \sigma \delta(\mathbf{y}) \mathrm{tr}\,( \tilde \Lambda \tilde A_{\uk} )$, where $\tilde A_{\uk}{}^\alpha$ is a localised gauge field\footnote{Actually, in some SSCs, corresponding to cases where the fixed point plane does not fill all of spacetime, the $\tilde A_{\uk}{}^\alpha$ correspond to adjoint scalars. This is determined by the choice of overall ExFT section condition solution, which can imply that some $\partial_{\uk} = 0$. In the type II orientifold case, for instance, they can be thought as the usual scalar fields in the low energy effective actions on D$p$ branes, T-dual to the Yang-Mills gauge fields in the type I theory.} carrying an ``internal'' index $\uk$ (this appears in the expansion of the half-maximal structure itself, or equivalently the generalised metric which at the fixed point is in fact expressed solely in terms of the half-maximal structure). Here $\sigma$ is a constant pre-factor, and the trace is in the adjoint of the gauge group $G$ ($\mathrm{tr} ( \tilde \Lambda_1 \tilde \Lambda_2 ) \equiv \kappa_{\alpha \beta} \tilde \Lambda_1^\alpha \tilde \Lambda_2^\beta$). We similarly expand the gauge parameter $\lambda_\mu{}^{MN}$ in terms of the generalised tensor $n$ appearing in \eqref{basis}.

Then, studying the expansion of \eqref{deltaA}, and using \eqref{modgld} as well as further compatibility conditions involving the basis \eqref{basis}, we find that that the components ${A}_{\mu \uk}$ pick up a localised gauge transformation under gauge transformations $\tilde \Lambda^\alpha$ of the extra gauge fields, of the form  
\be
\delta {A}_{\mu \uk} \propto \delta( \mathbf{y} ) \kappa_{\alpha \beta} \tilde \Lambda^\alpha ( D_\mu \tilde A_{\uk}{}^\beta - \partial_{\uk} \tilde A_\mu{}^\beta ) \,,
\label{gaugelocal} 
\ee
where $D_\mu \equiv \partial_\mu - L_{A_\mu}$ a covariantised external partial derivative, with $L_{A_{\mu}}$ the ordinary Lie derivative with respect to $A_\mu{}^{\uk}$. 

This modified gauge transformation can be interpreted in different SSCs.
Depending on the SSC, all, some or none of the $\mathbf{y}$ are physical directions. For instance,
\begin{itemize}
\item in the Ho\v{r}ava-Witten SSC, the full set of odd coordinates are $\mathbf{y} \equiv (Y^{12}, Y^{13}, Y^{14}, Y^{15} )$, while the physical coordinates are $Y^{i5}$, $i=1,\dots,4$.
So, the only odd physical coordinate is $Y^{15}$. Focusing solely on the fixed point at $Y^{15} = 0$, we take the delta function in \eqref{eta} to be just $\delta(\mathbf{y} ) \rightarrow \delta(Y^{15})$.
A close inspection of the dictionary relating ExFT to 11-dimensional SUGRA reveals that the components $A_{\mu \uk}$ can be identified with the even components $C_{\mu \uk 1}$ of the 11-dimensional three-form, which indeed receive localised modified gauge transformations in the Ho\v{r}ava-Witten theory.
\vspace{0.5em}

\item in a heterotic SSC, or that corresponding to type I, the full set of coordinates $\mathbf{y}$ are always dual coordinates only, so that we replace $\delta(\mathbf{y}) \rightarrow 1$.
Then the $A_{\mu \uk}$ correspond to components $B_{\mu \uk}$ of the 10-dimensional NSNS 2-form (in heterotic) or components $C_{\mu \uk}$ of the 10-dimensional RR 2-form (in type I), and the localised gauge transformation \eqref{gaugelocal} is the usual Green-Schwarz modification. 
\end{itemize}

This analysis can be extended to the other ExFT fields in order to obtain the full set of modified gauge transformations of the requisite form field in each SSC.
Continuing in this way, we reproduce the field content, extra gauge fields, modified Bianchi identities, full modified (Green-Schwarz) gauge transformations, and actions of all the theories in the half-maximal duality web.
This is explained at length in \cite{Blair:2018lbh}.

\section{Discussion and open problems}

We have outlined in this proceedings some of the features of O-folds, as can be studied in the context of exceptional field theory. 	
We saw that a simple $\mathbb{Z}_2$ quotient provided a unified description of type II orientifolds, heterotic and Ho\v{r}ava-Witten, and that we could include localised vector multiplets within ExFT. 
This was carried out in detail for the $\mathrm{SL}(5)$ ExFT, but the general idea will go through to other groups.

The outstanding open problem is to develop a method to determine the gauge groups appearing at fixed points.
For the $\mathbb{Z}_2$ quotient, this requires an understanding of anomaly cancellation in ExFT.
The hope would be that the unification of metric and gauge degrees of freedom in ExFT, plus the fact that it describes the trivially anomaly free (non-chiral) IIA theory on the same footing as the potentially anomalous (chiral) IIB theory, might be a simplifying factor. This would presumably require the introduction of concepts such as topological invariants in ExFT, which are not currently understood.

It would be interesting to consider the moduli space of O-fold compactifications, and explore gauge enhancement. Even for the $\mathbb{Z}_2$ O-fold alone this may be rich to explore. 
A first step here could be to look at the type I${}^\prime$ / heterotic SSCs and understand whether the heterotic DFT description of gauge enhancement of \cite{Aldazabal:2017jhp,Fraiman:2018ebo} can be embedded in ExFT. On the heterotic side, this corresponds to tuning Wilson lines and other moduli, while on the type I${}^\prime$ side this translates into statements about the locations of D-branes on the interval: in the ExFT O-fold picture, these should be different facets of a unified picture. Additionally, the M-theory geometric orbifold \eqref{MsecO6} can also be viewed as M-theory on an orbifold limit of K3, and we could imagine seeking to understand ExFT on the fully resolved K3, generalising the approach of \cite{Malek:2016vsh}.

The classification of possible O-folds, and whether they are geometric/non-geometric in different SSCs, for different $\mathrm{E}_{d(d)}$ groups and potentially different amounts of preserved supersymmetries, is an obvious avenue to pursue. Note that there can be new features, e.g. in the $d=5$ $\mathrm{SO}(5,5)$ there are two types of half-maximal structures (chiral and non-chiral) leading to a richer spectrum of quotiented theories \cite{Blair:2018lbh}.

Generically, an O-fold will be highly non-geometric, with physical coordinates $Y$ identified with dual coordinates $\tilde Y$ in all solutions of the section condition.
Whether or not one can obtain some amount of control over such configurations in ExFT remains to be seen. 

\section*{Acknowledgements}

I would like to thank the organisers of the Corfu Summer Institute 2018 workshop on ``Dualities and Generalised Geometries'' for inviting me to speak, and my collaborators Emanuel Malek and Dan Thompson for feedback on this manuscript. I am supported by an FWO-Vlaanderen Postdoctoral Fellowship, and in part by the FWO-Vlaanderen through the project G006119N and by the Vrije Universiteit Brussel through the Strategic Research Program ``High-Energy Physics''.

\bibliography{NewBib}

\end{document}